\documentclass[graybox]{svmult}

\usepackage{mathptmx}       % selects Times Roman as basic font
\usepackage{helvet}         % selects Helvetica as sans-serif font
\usepackage{courier}        % selects Courier as typewriter font
\usepackage{type1cm}        % activate if the above 3 fonts are
\usepackage{latexsym}
\usepackage{ams symb}
\usepackage{amsmath}
\usepackage{rotating}
\usepackage{makeidx}         % allows index generation
\usepackage{graphicx}        % standard LaTeX graphics tool
\usepackage{multicol}        % used for the two-column index
\usepackage[bottom]{footmisc}% places footnotes at page bottom

\makeindex             % used for the subject index
\begin{document}

\title*{Exploring the  Universe with Metal-Poor Stars}
\titlerunning{Exploring the Universe with Metal-Poor Stars}
\author{Anna Frebel}
\institute{Anna Frebel \at Harvard-Smithsonian Center for Astrophysics,
60 Garden St.,
Cambridge, MA 02138, USA, \\\email{afrebel@cfa.harvard.edu}}

\maketitle

\abstract{The early chemical evolution of the Galaxy and the Universe
is vital to our understanding of a host of astrophysical phenomena.
Since the most metal-poor Galactic stars (with metallicities down to
$\mbox{[Fe/H]}\sim-5.5$) are relics from the high-redshift Universe,
they probe the chemical and dynamical conditions of the Milky Way and
the origin and evolution of the elements through nucleosynthesis. They
also provide constraints on the nature of the first stars, their
associated supernovae and initial mass function, and early star and
galaxy formation.
The Milky Way's dwarf satellites contain a large fraction ($\sim30\%$)
of the known most metal-poor stars that have chemical abundances that
closely resemble those of equivalent halo stars. This suggests that
chemical evolution may be universal, at least at early times, and that
it is driven by massive, energetic SNe. Some of these surviving, ultra-faint
systems may show the signature of just one such Pop\,III star; they
may even be surviving first galaxies.
Early analogs of the surviving dwarfs may thus have played an
important role in the assembly of the old Galactic halo whose
formation can now be studied with stellar chemistry. Following the
cosmic evolution of small halos in simulations of structure formation
enables tracing the cosmological origin of the most metal-poor stars
in the halo and dwarf galaxies. Together with future observations and
additional modeling, many of these issues, including the reionization
history of the Milky Way, may be constrained this way.  The chapter
concludes with an outlook about upcoming observational challenges and
ways forward is to use metal-poor stars to constrain theoretical
studies.}

%%%%%%%%%%%%%%%%%%%%%%%%%%%%%%%%%%%%%%%%%%%%%%
\section{Introduction}

As Carl Sagan once remarked, \textit{If you wish to make an apple pie
from scratch, you must first create the Universe}. An apple contains
at least 16 different elements, and the human body is even more
complex, having at least trace amounts of nearly 30
elements\footnote{http://chemistry.about.com/cs/howthingswork/f/blbodyelements.htm},
all owing to a 14-billion year long manufacturing process called
cosmic chemical evolution.  Thus, the basis of chemically complex and
challenging undertakings such as cooking and baking, not to mention
the nature of life, will ultimately be gained through an understanding
of the formation of the elements that comprise organic material. It is
thus important to examine how the constituents of an apple, and by
extension the stuff of life and the visible Universe were created:
baryonic matter in the form of elements heavier than primordial
hydrogen and helium.  

\begin{figure*}  [!t]
\begin{center}
\includegraphics[width=11.7cm,clip=true,bbllx=65,bblly=423,bburx=528,bbury=655]{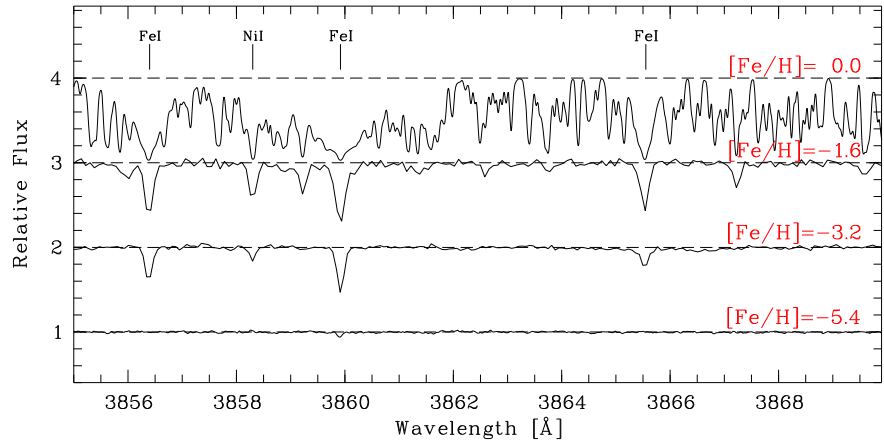}
  \caption{\label{spec_comp} Spectral comparison of stars in the
  main-sequence turn-off region with different metallicities. Several
  absorption lines are marked. The variations in line strength reflect
  the different metallicities. From top to bottom: Sun with
  $\mbox{[Fe/H]}=0.0$, G66-30 $\mbox{[Fe/H]}=-1.6$ (Norris et al. 1997)
  G64-12 $\mbox{[Fe/H]}=-3.2$ (Aoki et al. 2006), and HE1327-2326
  $\mbox{[Fe/H]}=-5.4$ (Frebel et al. 2005).  Figure taken from
  Frebel (2010).}
\end{center}
  \end{figure*}

This chapter aims at describing that the chemical abundances observed
in the most metal-poor stars can be employed to unravel a variety of
details about the young Universe, such as early star formation
environments, supernovae (SNe) nucleosynthesis, and the formation
process(es) of the Galactic halo. To illustrate the meaning of
low-metallicity, Figure~\ref{spec_comp} shows the progression from
metal-rich to the most metal-poor stars; spectra around the strongest
optical Fe line at 3860\,{\AA} are shown of the Sun and three
metal-poor main-sequence turn-off stars. The number of atomic
absorption lines detectable in the spectra decreases with increasing
metal-deficiency. In HE~1327$-$2326, the star with the currently low
Fe abundance, only the intrinsically strongest metal lines remain
observable. As can be seen in the Figure, these are extremely weak. If
a main-sequence star with even lower Fe value (or somewhat hotter
temperature) was discovered, no Fe lines would be measurable
anymore. In the case of a metal-deficient giant, the lines would be
somewhat stronger due to its cooler temperature and thus allow for the
discovery of an object with $\mbox{[Fe/H]}\lesssim-6$.

Because these most metal-poor stars represent easily accessible
\textit{local equivalents} of the high-redshift universe, and as such,
provide a unique tool to address a wide range of near and far-field
cosmological topics. In short, metal-poor stars enable scientific
progress in three areas that bridge our understanding of the current
state of the Galactic halo and its old stellar population with that of
the evolution of local dwarf galaxies, to the formation of large
galaxies like the Milky Way more generally, as well as the beginning
of star and galaxy formation in the early universe.

\begin{itemize}

\item[1.] \textit{Stellar Archaeology:} Constrains the astrophysical sites and
conditions of nucleosynthesis and the major physical processes that
drove early star formation. Abundance measurements of many elements
throughout the periodic table of metal-poor halo stars enable the
detailed documentation of the earliest chemical enrichment events.

\item[2.] \textit{Dwarf Archaeology:} Provides constraints on galaxy
formation on small scales, and metal mixing and feedback processes. By
comparing abundances of metal-poor stars in ultra-faint dwarf galaxies
to those of equivalent halo stars, the universality of the (beginning
of) chemical evolution can be tested, what the relation is between the
dwarfs and the ``building blocks'' of the Galactic halo, and whether
they are the survivors of the first galaxies.

\item[3.] \textit{Near-Field Cosmology:} Determines the role of
metal-poor stars as tracers of the accretion history of the Milky Way
halo. The coupling of low-metallicity stellar abundances with results
from cosmological simulations enables the study of the formation
mechanism(s) of large galaxies like the Milky Way with its old halo
and satellites.

\end{itemize}

For stellar archaeology, large numbers of Galactic metal-poor halo
stars, mostly found in objective-prism surveys in both hemispheres
such as the HK survey of Beers and collaborators, the Hamburg/ESO
survey (Christlieb and collaborators), and more recently SDSS, are
needed to gain detailed insights into the history and evolution of our
Galaxy (e.g., {Beers} \& {Christlieb} 2005; Frebel \& Norris 2011). For
dwarf archaeology, observations of any stars in dwarf satellite
galaxies orbiting the Milky Way are required, although these are more
difficult to obtain than those of the halo stars. Finally, near-field
cosmology encourages the systematic use of metal-poor stars for
studying galaxy formation and cosmological aspects.  In a universe
dominated by cold dark matter (CDM), like the one we live in, galaxy
formation proceeds hierarchically through the accretion of smaller
objects onto the main halo. Simulations show that successive growth is
reflected in the abundance of dark matter substructures in the halos
of large galaxies like the Milky Way, and it is believed that the
luminous satellites of our Galaxy are the visible counterparts to at
least some of these substructures. Thus, the collective body of
metal-poor stars now found in the halo as well as the dwarf galaxies
enable addressing a number of important, outstanding questions that
show how closely connected the three topics are.

%-------------------------------
\begin{itemize}

\item What is the nature of Pop\,III stars? Are the yields of
  the first SNe different from today's? Can the
  signatures of theorized pair-instability SNe be found in metal-poor stars?
\item What drove early star formation?  How/where did the first
low-mass stars and the first galaxies form?
\item What are the main nucleosynthesis processes and sites that are
  responsible for forming the elements from the Big Bang until today?
\item How did chemical evolution proceed? How do stellar chemistry and
 halo kinematics correlate? How can the abundances be used to learn
 about the halo formation process?
\item Was the old halo built from accreted satellites? Can accreted
dwarf galaxies be identified in the halo? Did the first stars form in
dwarf galaxies?
\end{itemize}

Each section of this chapter discusses a significant aspect in which
metal-poor stars offer unique insight into the young
universe. Section~2 sets the overall stage by introducing the first
stars, the halo metallicity distribution function and considerations
regarding early low-mass star formation. Section~3 describes the role
of metal-poor stars in the Galactic as tracers of the earliest
enrichment events and chemical evolution. This concept is extended to
dwarf galaxies in Section~4. Understanding the formation history of
the Milky Way with the help of metal-poor stars is outlined in
Section~5.  Conclusions and an outlook at given in Section~6.

\section*{Further Reading \& Definitions}

This chapter is partially based on an article by
Frebel (2010)\footnote{The 2009 Biermann Award Lecture, originally
published in Astronomische Nachrichten by VCH/Wiley, 2010, 331, 474}
that, among other topics, introduces metal-poor stars as probes for
theoretical works related to the early universe. Here, this discussion
is extended to showcase the versatility and potential of the oldest
stars for dwarf archaeology and near-field cosmology. The main aim is
to outline the broad picture of studying galaxy formation processes
with stellar chemistry. For a more in-depth discussion of stellar
abundances, abundance derivations, details on nucleosynthesis and
chemical evolution, kinematics, stellar age determinations and
cosmogony, the reader is referred to Frebel \& Norris (2011) and references
therein.

Since there exist a large range of metal-poor stars in terms of their
metallicities and chemical signatures, {Beers} \& {Christlieb} (2005) suggested a
classification scheme. Extensive use will be made of their term
``extremely metal-poor stars'', referring to stars with
$\mbox{[Fe/H]}<-3.0$.  This nomenclature shows that the main
metallicity indicator used to determine any stellar metallicity is the
iron abundance, [Fe/H], which is defined as \mbox{[A/B]}$ =
\log_{10}(N_{\rm A}/N_{\rm B})_\star - \log_{10}(N_{\rm A}/N_{\rm
B})_\odot$ for the number N of atoms of elements A and B, and $\odot$
refers to the Sun. For example, $\mbox{[Fe/H]}=-3.0$ is 1/1000 of
solar Fe abundance.  With few exceptions, [Fe/H] traces the overall
metallicity of the objects fairly well.

\section{Exploring The Early Universe with Metal-Poor Stars}

\subsection{The First Stars}

According to cosmological simulations that are based on the $\Lambda$
cold dark matter model of hierarchical structure growth in the
Universe, the first stars formed in small minihalos some few hundred
million years after the Big Bang. Due to the lack of cooling agents in
the primordial gas, significant fragmentation was largely suppressed
so that these first objects were very massive (of the order to
$\sim100 $\,M$_{\odot}$; e.g, {Bromm} \& {Yoshida} (2011) and references
therein). This is in contrast to low-mass stars dominating today's
mass function. These objects are referred to as Population\,III
(Pop\,III) as they formed from metal-free gas. Recent modeling of
first star formation suggests that these early behemoth were rapidly
rotating (Stacy et al. 2010) and new observations have provided evidence in
support of this claim (Chiappini et al. 2011). Moreover, significant
fragmentation of the star forming cloud may occur that could lead to
multiple first stars in a given minihalo (Clark et al. 2011).

The stars soon exploded as SNe to either collapse into black holes
(progenitor masses of $25<M_{\odot}<140$ and $M_{\odot}>260$) or to
die as energetic pair-instability SNe (PISN; $140<M_{\odot}<260$;
Heger et al. 2002). During their deaths, these objects provided vast
amounts of ionizing radiation (and some of the first metals in the
case of the PISNe) that changed the conditions of the surrounding
material for subsequent star formation even in neighboring minihalos.
Hence, the second generation of stars might have been less massive
(M$_{\star}\sim10\,$M$_{\odot}$). Partially ionized gas supports the
formation of the H$_{2}$, and then the HD molecule, which in turn
facilitates more effective cooling than what is possible in neutral
gas. Also, any metals or dust grains left behind from PISNe would have
similar cooling effects. This may then have led to the first more
regular metal-producing SNe, although not all higher mass SNe must
necessarily end in black hole formation. Umeda \& Nomoto (2003)
suggested that some 25\,M$_{\odot}$ stars undergo only a partial
fallback, so that only some of the newly created metals get ejected
into the surrounding gas.

By that time, most likely enough metals were present to ensure
sufficient gas fragmentation to allow for low-mass ($<$1\,M$_{\odot}$)
star formation.  Stars that formed from any metal-enriched material
are referred to as Population\,II (Pop\,II) stars. More metal-rich
stars like the Sun that formed in a much more metal-rich Universe are
called Population\,I. Studying the ``chemical fingerprints'' of the
oldest, most metal-poor stars (extreme Pop\,II) reveals information
about the first nucleosynthesis events in the Universe; indeed,
several metal-poor star abundance patterns have been fitted with
calculated Pop\,III SN yields (see Section~\ref{mmps}). Moreover,
evidence for the existence of PISNe could potentially be obtained if
their characteristic signature (a pronounced effect in the abundance
signature of elements with odd or even atomic number) were found in
metal-poor stars. This has, however, not yet occurred.

\subsection{The Metallicity Distribution Function of the Galactic Halo}

The metallicity distribution function (MDF) represents the integrated
chemical evolution of a system which began with the first stars and
was continued by many stellar generations at various astrophysical
sites and over different timescales.

\begin{figure*}[!t]
\center{
\includegraphics[width=0.82\textwidth,clip=true,bbllx=80,bblly=38,bburx=550,bbury=775]{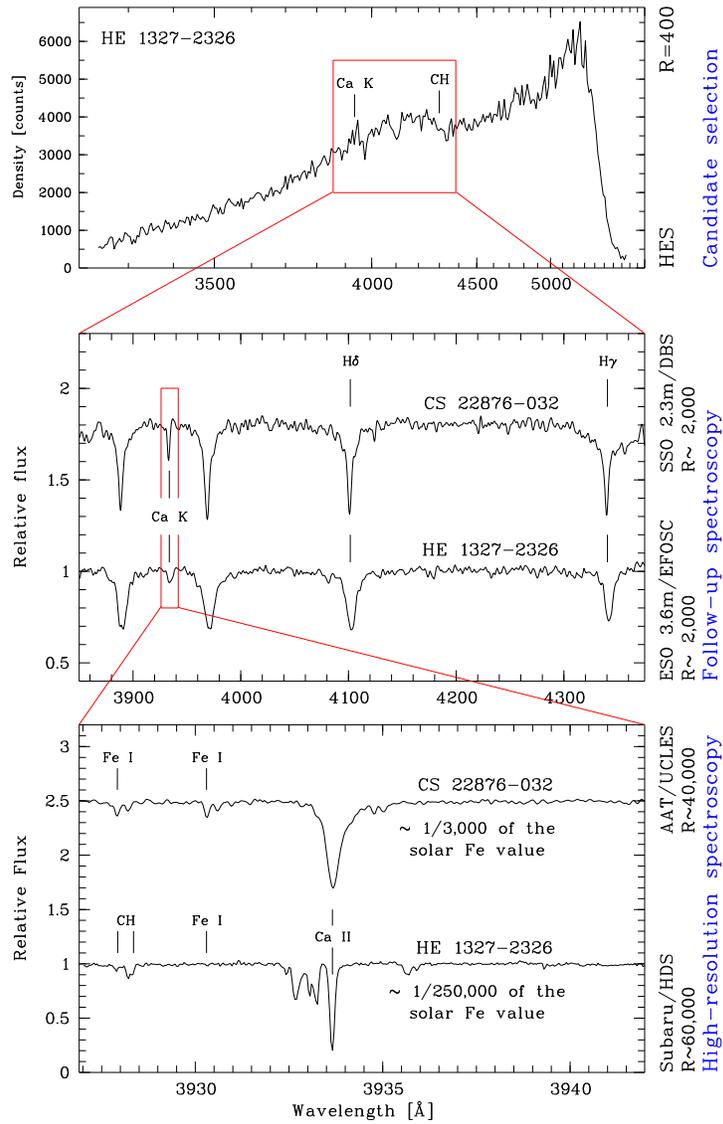}
}
\caption{\label{fig:3steps} The three observational steps to find
  metal-poor stars illustrated by means of HE~1327$-$2326. Top panel:
  HES objective-prism spectrum. Middle panel: Medium-resolution
  spectrum of HE~1327$-$2326 in comparison with CS~22876$-$032
  ($\mbox{[Fe/H]}=-3.7$; Norris et al. 2000 and references
  therein). From this data we measured $\mbox{[Fe/H]}=-4.3$ for
  HE~1327$-$2326 because interstellar Ca blended with the Ca II K
  line. Bottom panel: High-resolution spectra of both objects. Only
  with the high-resolution data was it possible to determine the true
  iron abundance, $\mbox{[Fe/H]}=-5.4$, for HE~1327$-$2326. Figure
  taken from Frebel et al. (2005b).}
\end{figure*}

To establish the MDF of a given system, a large, complete sample of
stars with good metallicity estimates is required. Over the past two
decades, the quest to find the most metal-poor stars to study the
chemical evolution of the Galaxy led to a significant number of stars
with metallicities down to $\mbox{[Fe/H]}\sim-4.0$ (see {Beers} \&
{Christlieb} 2005 for a more detailed review). Those stars were
initially selected as candidates from a large survey, such as the HK
survey (Beers et al. 1992) and the Hamburg/ESO survey (Wisotzki et
al. 1996).  A large survey is required to provide numerous
low-resolution spectra to search for weak-lined stellar candidates
indicating metal deficiency. Those spectra have to cover the strong
Ca\,II\,K line at 3933\,{\AA} because the strength of this line
indicates the metallicity of the star, and can be measured even in
low-quality spectra. This is shown in Figure~\ref{fig:3steps}.  If
this line is sufficiently weak as a function of the star's estimated
effective temperature, an object is selected as a candidate metal-poor
star. For all candidates, medium-resolution spectra ($R\sim2,000$) are
required to more accurately determine the Ca\,II\,K line strength for
a more robust estimate for the Fe abundance. This line is still the
best indicator for the overall metallicity [Fe/H] of a metal-poor star
in such spectra. In the Sloan Digital Sky Survey and LAMOST survey,
the survey spectra themselves are already of medium-resolution,
allowing for a quicker and more direct search for metal-poor
stars. Photometric surveys like Skymapper with extensive filter sets
designed for stellar work (Keller et al. 2007) will also yield large
numbers of high-quality candidates.

%----------------------------------------------
%----------------------------------------------

To confirm the metallicity, and to measure elemental abundances from
their respective absorption lines besides that of iron,
high-resolution ($R>20,000$) optical spectroscopy is required (see
bottom panel of Figure~\ref{fig:3steps}). Only then the various
elements become accessible for studying the chemical evolution of the
Galaxy. Those elements include carbon, magnesium, calcium, titanium,
nickel, strontium, and barium, and trace different enrichment
mechanisms, events and timescales. Abundance ratios [X/Fe] as a
function of [Fe/H] can then be derived for the lighter elements
($Z<30$) and neutron-capture elements ($Z>38$). The resulting
abundance trends will be further described in Section~3.2. The final
number of elements thereby depends on the type of metal-poor star, the
wavelength coverage of the data, and the data quality itself.

\begin{figure}  [!htb]
\includegraphics[width=11.7cm,clip=true,bbllx=40,bblly=440,bburx=470,bbury=752]{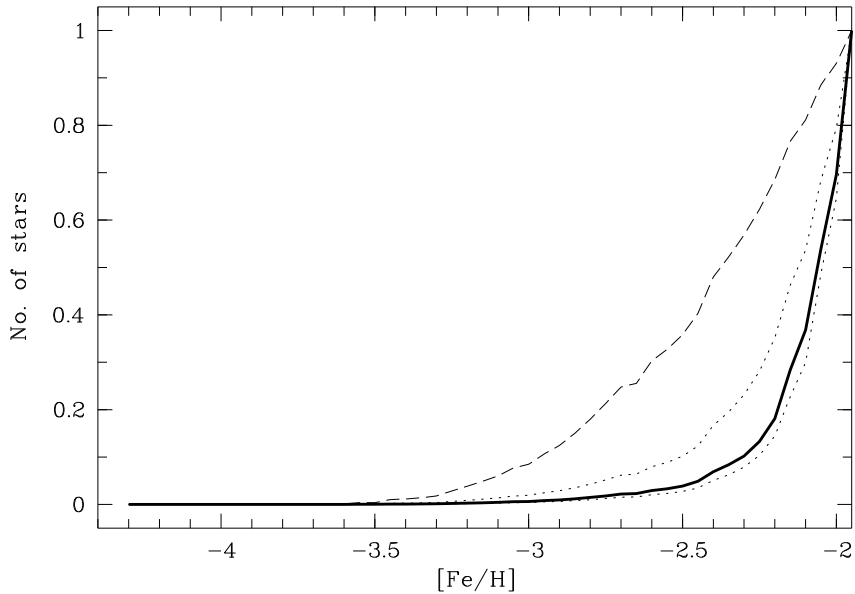}
  \caption{ \label{mdf} Cumulative metallicity distribution function
  of the Galactic halo based on metal-poor giants selected form the
  Hamburg/ESO survey (Sch\"oerck et al. 2009). The solid line represents the
  bias-corrected MDF, whereas the dotted lines show the level of
  uncertainty in the MDF based on the various correction
  functions. The dashed line shows the uncorrected, ``as observed''
  MDF -- The difference to the solid line shows how successful
  searches for the most metal-poor stars have been. }
\end{figure}

Sch{\"o}rck et al. (2009) and Li et al. (2010) presented MDFs for halo stars that
are corrected for various selection effects and other biases. The
number of known metal-poor stars declines significantly with
decreasing metallicity (below $\mbox{[Fe/H]}<-2.0$) as illustrated in
Figure~\ref{mdf}. Only very few stars are known ($\lesssim30$) with
metallicities below $\mbox{[Fe/H]}<-3.5$, but it is these objects that
enable the most insight into the early universe and the beginning of
chemical evolution.

The bias-corrected MDF shows how rare metal-poor stars really are, but
also, that past targeted (``biased'') searches for metal-poor stars
have been extremely successful at identifying these rare objects
(e.g., Frebel et al. 2006; {Christlieb} {et~al. }2008). The most important
achievements in terms of the most iron-deficient stars was the push to
a significantly lower stellar metallicity $\mbox{[Fe/H]}$ almost a
decade ago: From a longstanding $\mbox{[Fe/H]}=-4.0$ (CD $-38^{\circ}$
245; Bessell \& Norris 1984) to $\mbox{[Fe/H]}=-5.2$ (HE~0107$-$5240;
Chrislieb et al. 2002)\footnote{Applying the same non-LTE correction to
the Fe\,I abundances of HE~0107$-$5240 and HE~1327$-$2326 leads to a
final abundance of $\mbox{[Fe/H]}=-5.2$ for HE~0107$-$5240.}), and
down to $\mbox{[Fe/H]}=-5.4$ more recently (HE~1327$-$2326;
Frebel et al. 2005). Overall, only three stars are known with iron
abundances of $\mbox{[Fe/H]}<-4.0$. The third star, HE~0557$-$4840
Norris et al. 2007 with $\mbox{[Fe/H]}<-4.8$, bridges the gap between
$\mbox{[Fe/H]}=-4.0$ and the two hyper Fe-poor objects.

Beyond these individual stars that form the very tail of the MDF, a
number of studies involving samples of $\sim$10-30 stars in the
metallicity range of $-4.0\mbox{[Fe/H]}<-2.5$ have been carried out
over the years. These works have delivered many important details
about chemical enrichment and nucleosynthesis, and greatly improved our
understanding of the early universe (e.g., 
McWilliam et al. 1995,
Ryan et al. 1996, 
Cayrel et al. 2004, 
Barklem et al. 2005,
Aoki et al. 2008,
Cohen et al. 2008, 
Lai et al. 2008, 
Hollek et al. 2011).

%, \cite{frebel}).

%norris+yong new

\subsection{Early low-mass star formation and the connection to 
carbon-enhanced metal-poor stars}\label{sec_arch}

Early Pop\,II stars began to form from the enriched material left
behind by the first stars. The actual formation process(es) of these
initial low-mass (M $\le$ 0.8 M$_{\odot}$) Pop\,II stars (i.e. the
most metal-poor stars) that live longer than a Hubble time, are,
however, not well understood so far. Ideas for the required cooling
processes necessary to induce sufficient fragmentation of the
near-primordial gas include cooling through metal enrichment
(``critical metallicity'') or dust, cooling based on enhanced molecule
formation due to ionization of the gas, as well as more complex
effects such as turbulence and magnetic fields (Bromm et al. 2009).

Fine-structure line cooling through neutral carbon and singly-ionized
oxygen has been suggested as a main cooling agent facilitating
low-mass star formation (Bromm \& Loeb 2003).  These elements were
likely produced in vast quantities in Pop\,III objects (e.g. Meynet et
al. 2006; Chiappini et al. 2011). Gas fragmentation is then induced
once a critical metallicity of the interstellar medium (ISM) is
achieved.  The existence of such a critical metallicity can be probed
with large numbers of carbon and oxygen-\textit{poor} metal-poor
stars.  Frebel et al. (2007b) developed an ``observer-friendly''
description of the critical metallicity that incorporates the observed
C and/or O stellar abundances; $D_{\rm trans}={\rm log} (10^{{\rm
[C/H]}} + 0.3 \times 10^{{\rm [O/H]}})\ge-3.5$.  Any low-mass stars
still observable today then has to have C and/or O abundances above
the threshold of $D_{\rm trans}=-3.5$ (see Figure~1 in
Frebel et al. 2007). At metallicities of $\mbox{[Fe/H]}\gtrsim-3.5$, most
stars have C and/or O abundances that are above the threshold since
they follow the solar C and O abundances simply scaled down to their
respective Fe values. Naturally, this metallicity range is not
suitable for directly probing the first low-mass stars. Below
$\mbox{[Fe/H]}\sim-3.5$, however, the observed C and/or O levels must
be higher than the Fe-scaled solar abundances to be above the critical
metallicity. Indeed, none of the known lowest-metallicity stars has a
$D_{trans}$ below the critical value, consistent with this cooling
theory. Some stars, however, have values very close to $D_{\rm
trans}=-3.5$.  HE~0557$-$4840, at $\mbox{[Fe/H]}=-4.75$ (Noris et al. 2007),
falls just onto the critical limit (M. Bessell 2009, priv.  comm.). A
star in the ultra-faint dwarf galaxy Bo\"otes\,I has $D_{\rm
trans}=-3.2$ (at $\mbox{[Fe/H]}=-3.7$; and assuming that its oxygen
abundance is twice that of carbon). Another interesting case is the
most metal-poor star in the classical dwarf galaxy Sculptor, which has
an upper limit of carbon of $\mbox{[C/H]}<-3.6$ at
$\mbox{[Fe/H]}=-3.8$ (Frebel et al. 2010b). Despite some still required
up-correction of carbon to account for atmospheric carbon-depletion of
this cool giant, the star could potentially posses a sub-critical
$D_{\rm trans}$ value.

Overall, more such ``borderline'' examples are crucial to test for the
existence of a critical metallicity. If fine-structure line cooling
were the dominant process for low-mass star formation, two important
consequences would follow: 1) Future stars to be found with
$\mbox{[Fe/H]}\lesssim-4.0$ are predicted to have these substantial C
and/or O overabundances with respect to Fe.  2) The so-far unexplained
large fraction of metal-poor objects that have large overabundances of
carbon with respect to iron ($\mbox{[C/Fe]}>1.0$) may reflect an
important physical cause.  About 20\% of metal-poor stars with
$\mbox{Fe/H}\lesssim-2.5$ exhibit this behavior
(e.g. {Beers} \& {Christlieb} 2005). Moreover, at the lowest metallicities, this
fraction is even higher. All three stars with $\mbox{[Fe/H]}<-4.0$ are
extremely C-rich, well in line with the prediction of the line cooling
theory.

This may, however, not the only way for forming low-mass stars.
Cooling through dust grains might also have been responsible for the
transition from Pop\,III to Pop\,II star formation. Dust created in
high-density regions during the first SNe explosions or mass loss
during the evolution of Pop\,III stars may induce fragmentation
processes (e.g., Schneider et al. 2006) that lead to the formation of
subsolar-mass stars. The critical metallicity in this scenario is a
few orders of magnitude below that of C and O line cooling.  If some
metal-poor stars are found to be significantly below $D_{\rm
trans}=-3.5$, their existence may still be consistent with the
critical value set by dust cooling. Irrespective of the differences in
cooling channels, such criteria will need to be incorporated in
large-scale simulations to take environmental influences, such as the
available gas mass, into account.

%%%%%%%%%%%%%%%%%%%%%%%%%%%%%%%%%%%%%%%%%%%%%%
\section{Stellar Archaeology}

\subsection{Validating Stellar Archaeology}

The concept of stellar archaeology is based on long-lived low-mass
metal-poor main-sequence and giant stars whose chemical abundances are
thought to reflect the composition of the gas cloud during their
formation period. A vital assumption is that the stellar surface
compositions have not been significantly altered by any internal
mixing processes given that these stars are fairly unevolved despite
their old age. But are there other means by which the surface
composition could be modified?  Accretion of interstellar matter while
a star orbits in the Galaxy for $\sim10$\,Gyr has long been suggested
as a mechanism to affect the observed abundance
patterns. Iben (1983) calculated a basic "pollution limit" of
$\mbox{[Fe/H]}=-5.7$ based on Bondi-Hoyle accretion. He predicted that
no stars with Fe abundances below this value could be identified as
such since they would have accreted too much enriched material.

Assuming that stars with such low-metallicities exist (for example
low-mass Pop\,III stars if the IMF was Salpeter-like, and not
top-heavy), significant amounts of interstellar accretion could
masquerade the primordial abundances of those putative low-mass
Pop\,III stars. Analogously, stars with very low abundances, say
$\mbox{[Fe/H]}<-5.0$, could principally be affected also. To assess
the potential accretion level, Frebel et al. (2009) carried out a kinematic
analysis of a sample of metal-poor stars to assess their potential
accretion histories over the past 10\,Gyr in a Milky Way-like
potential. The amount of accreted Fe was calculated based on the total
accreted material over 10\,Gyr. The overall chemical evolution with
time was taken into account assuming the ISM to have scaled solar
abundances. The stellar abundances were found to be little affected by
accretion given their generally fast space velocities. The calculated,
``accreted abundances'' were often lower than the observed
measurements by several orders of magnitude. Johnson et al. (2011),
on the other hand, investigated direct accretion onto primordial
low-mass stars. If these stars had a weak solar-like wind it would
prevent the accretion of any material, at least in the early
universe. This would also be true for any low-metallicity stars,
although it was not considered by Frebel et al. (2009).

Generally, these studies show that accretion does not significantly
alter the observed abundance patterns, even in an extreme case in
which a star moves once through a very large, dense cloud.  The
concept of stellar archaeology can thus be deemed
viable. Nevertheless, since there is a large accretion dependency on
the space velocity it becomes obvious that kinematic information is
vital for the identification of the lowest-metallicity stars in the
Milky Way and the interpretation of their abundances. A way forward
would be an extensive assessment of potential gas accretion histories
for stars throughout the hierarchical assembly of a large
galaxy. However, the uncertainties regarding the existence and extent
of stellar winds may prevent strong conclusions.

\begin{figure}  [!t]
\centering
\includegraphics[width=8.7cm,clip=true,bbllx=43,bblly=40,bburx=550,bbury=752]{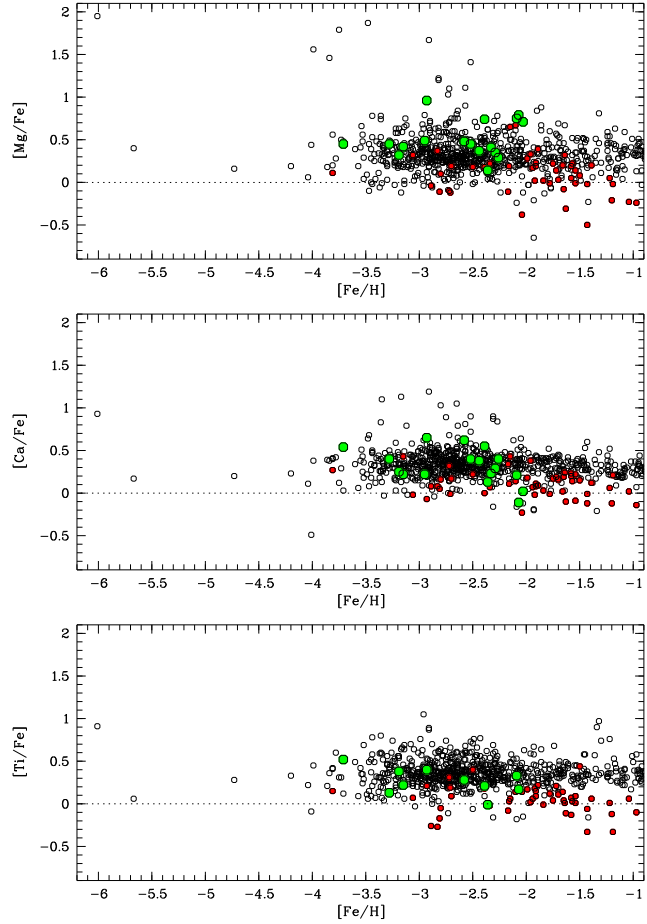}
  \caption{ \label{alphas} Light element abundance trends of Mg, Ca,
  and Ti. Black open circles represent halo stars, red filled circles
  are stars in the classical dwarf galaxies, and green filled circles
  show stars in the ultra-faint dwarf galaxies. The abundance data can
  be found in an electronic format. The scatter in the data likely
  reflects systematic differences between literature studies. Assuming
  this, systematic uncertainties in abundance analyses may be around
  $\sim0.3$\,dex. Figure taken from Frebel 2010.}
\end{figure}                                                            

\subsection{Chemical Evolution of the Galaxy}

Generally, there are several main groups of elements observed in
metal-poor stars, with each group having a common, main production
mechanism; 1) $\alpha$-elements (e.g. Mg, Ca, Ti) are produced
through $\alpha$-capture during various burning stages of late stellar
evolution, before and during SN explosions. These yields appear very
robust with respect to parameters such as mass and explosion energy 2)
Fe-peak elements (23$<Z<30$) are synthesized in a host of different
nucleosynthesis processes before and during SN explosions such as
radioactive decay of heavier nuclei or direct synthesis in explosive
burning stages, neutron-capture onto lower-mass Fe-peak elements
during helium and later burning stages and $\alpha$-rich freeze-out
processes. Their yields also depend on the explosion energy; 3) Light
and heavy neutron-capture elements ($Z>38$) are either produced in the
slow (s-) process occurring in thermally pulsing AGB stars (and then
transferred to binary companions or deposited into the ISM through
stellar winds) or in the rapid (r-) process most likely occurring in
core-collapse SN explosions. For more details on SN nucleosynthesis
see e.g., Woosley \& Weaver (1995).

The $\alpha$-element abundances in metal-poor halo stars with
$\mbox{[Fe/H]}<-1.5$ are enhanced by $\sim0.4$\,dex with respect to Fe
as seen in Figure~\ref{alphas}. This reflects a typical core-collapse
SN signature because at later times (in chemical space at about
$\mbox{[Fe/H]}\sim-1.5$) the onset of SN Ia provides a significant
contribution to the overall Galactic Fe inventory. As a consequence,
the \mbox{[$\alpha$/Fe]} ratio decreases down to the solar value at
$\mbox{[Fe/H]}\sim0.0$.  The general uniformity of light element
abundance trends down to $\mbox{[Fe/H]}\sim-4.0$ led to the conclusion
that the ISM must have been already well-mixed at very early times
(Cayrel et al. 2004). Otherwise it would be hard to understand why so
many of the most metal-poor stars have almost identical abundance
patterns. However, despite the well-defined abundance trends, some
stars, particularly those in the lowest metallicity regime show
significant deviations. Among those are some stars with unusually high
or low $\alpha$-element abundances.

\begin{figure}  [!t]
\centering
\includegraphics[width=8.7cm,clip=true,bbllx=43,bblly=40,bburx=550,bbury=752]{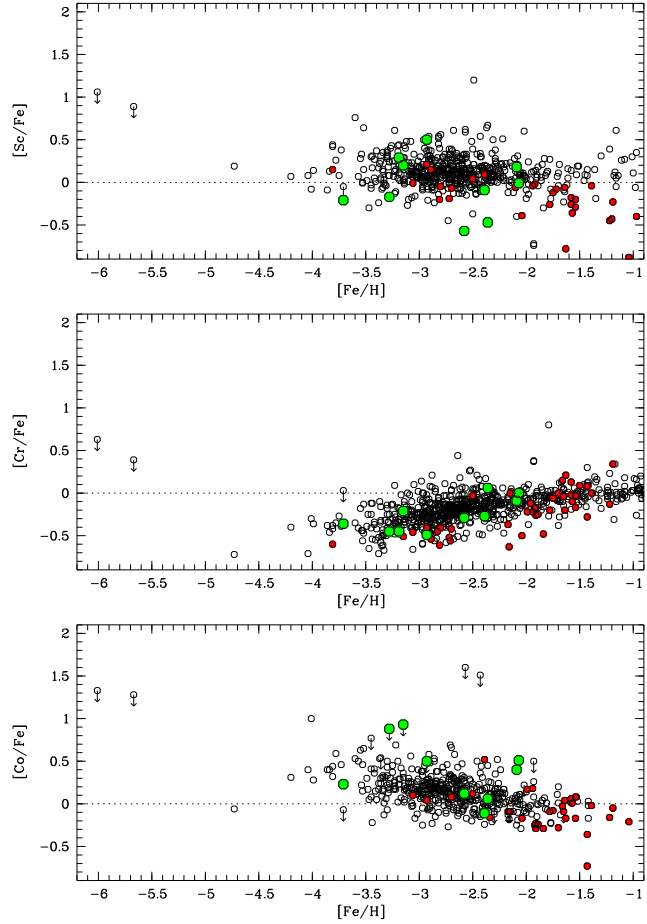}
  \caption{ \label{fepeak} Same as Figure~\ref{alphas}, but for the
  Fe-peak elements Sc, Cr, and Co. Figure taken from Frebel 2010.}
  \end{figure}

Among the Fe-peak elements, many have subsolar abundance trends at low
metallicity (e.g. [Cr,Mn/Fe]) which become solar-like as the
metallicity increases. This is shown in Figure~\ref{fepeak}. It is
not clear whether these large underabundances are of cosmic origin or
have to be attributed to modeling effects such as that of non-LTE
(Sobeck et al. 2007; Bergemann et al. 2008). Trends of other elements are somewhat
overabundant at low metallicity (Co) or relatively unchanged
throughout (Sc, Ni).  All elements with $Z<30$ hereby have relatively
tight abundance trends.

\begin{sidewaysfigure}
\begin{center}
\scalebox{0.5}
{\includegraphics[height=14.4cm]{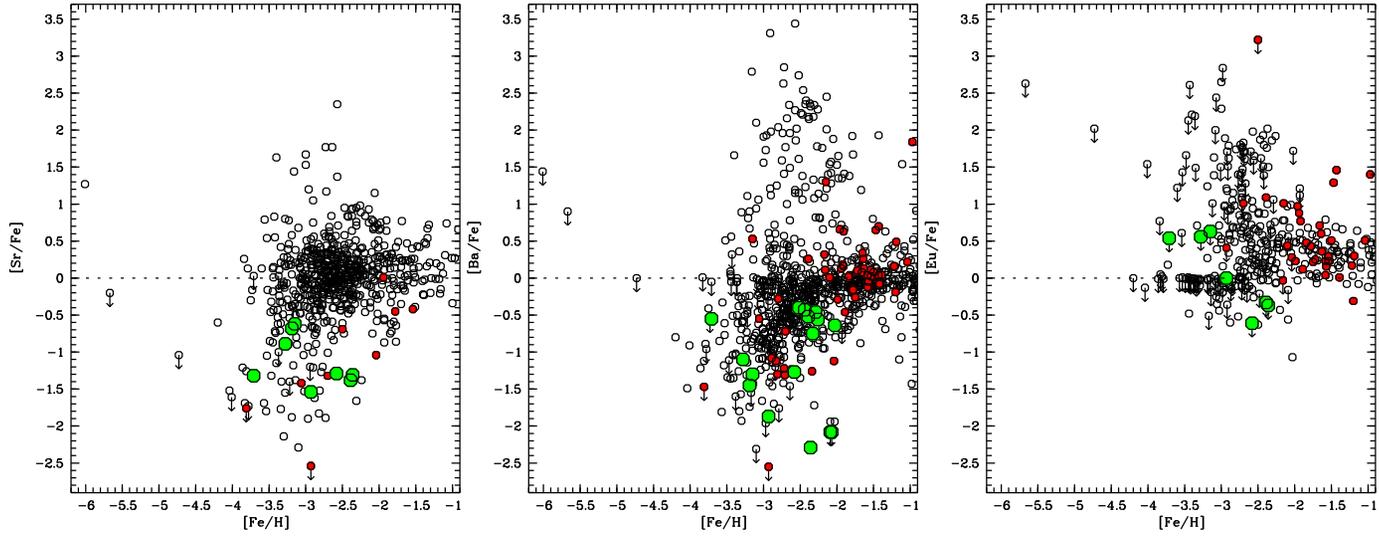}}
  \caption{ \label{ncap} Same as Figure~\ref{alphas}, but for the
  neutron-capture elements Sr, Ba, and Eu. For these elements, the
  scatter is much beyond systematic differences between individual
  studies and thus indicates a cosmic origin. Figure taken from
  Frebel 2010.}
\end{center}
\end{sidewaysfigure}

On the contrary, the abundances of the neutron-capture elements in
metal-poor stars are ``all over the place''.  Sr has an extremely
large scatter ($\sim 3$\,dex). This indicates that different
nucleosynthetic processes must have contributed to its Galactic
inventory, or that neutron-capture yields are very
environmentally-sensitive. Ba has even more scatter at
$\mbox{[Fe/H]}\sim-3.0$ (see Figure~\ref{ncap}). Some of this spread
may be explained through the existence of massive rotating
low-metallicity stars that produce large amounts of s-process like Sr
and Ba in the early universe (Chiappini et al. 2011). Other heavier
neutron-capture elements, such as r-process element Eu, have somewhat
less scatter. At the lowest metallicities, core-collapse SNe must have
dominated the chemical evolution (below $\mbox{[Fe/H]}=-3.0$). Hence,
the r-process is likely responsible for the neutron-capture elements
at this early time. Metal enrichment through the s-process began to
significantly contribute at somewhat later times, driven by the
evolutionary timescales of stars with $\sim2$-8\,M$_{\odot}$ to become
AGB stars. According to Simmerer et al. (2004), the s-process was in full
operation at $\mbox{[Fe/H]}\sim-2.6$.

\subsection{Tracing individual SN explosions with the 
most iron-poor stars}\label{mmps}

The faint ($V=15.2$) red giant HE~0107$-$5240 has $\mbox{[Fe/H]}=-5.2$
(Christlieb et al. 2002). The brighter ($V=13.5$) subgiant HE~1327$-$2326
has an even lower iron abundance of $\mbox{[Fe/H]}=-5.4$
(Frebel et al. 2005; Aoki et al. 2006). The latter value corresponds to
$\sim1/250,000$ of the solar iron abundance\footnote{Interestingly,
the entire mass of iron in HE~1327$-$2326 is actually 100 times less
than that in the Earth's core. At the same time the star is of course
of the order of a million times more massive than the Earth.}  A third
star with $\mbox{[Fe/H]}=-4.75$ (Norris et al. 2007) was found in 2006. The
metallicity of the giant HE~0557$-$4840 is in between the two
$\mbox{[Fe/H]}<-5.0$ stars and the next most metal-poor stars are
$\mbox{[Fe/H]}\sim-4.0$. Hence, it sits right in the previously
claimed ``metallicity gap'' (between $\mbox{[Fe/H]}\sim-4.0$ and
$\mbox{[Fe/H]}\sim-5.0$; e.g. Shigeyama et al. 2003) showing that the
scarcity of stars below $\mbox{[Fe/H]}=-4.0$ has no physical cause but
is merely an observational incompleteness. All three objects were
found in the Hamburg/ESO survey making it the so far most successful
database for metal-poor stars.

The most striking features in both $\mbox{[Fe/H]}<-5.0$ stars are the
extremely large overabundances of CNO elements.  HE~0557$-$4840 partly
shares this signature by also having a fairly large [C/Fe] ratio.
Other elemental ratios [X/Fe] are somewhat enhanced in HE~1327$-$2327
with respect to the stars with $-4.0<\mbox{[Fe/H]}<-2.5$, but less so
for the two giants.  Despite expectations, Li could not be detected in
the relatively unevolved subgiant HE~1327$-$2326.\footnote{The other
stars are giants. Thus, the surface Li is already destroyed due to the
thick convection zone transporting Li to deeper, hotter layers where
it burns.} The upper limit is $\log\epsilon ({\rm Li})<1.6$, where
$\log\epsilon ({\rm A})$ = $\log_{10}(N_{\rm A}/N_{\rm H})$ + 12. This
is surprising, given that the primordial Li abundance is often
inferred from similarly unevolved metal-poor stars
(Ryan et al. 1999). Furthermore, the upper limit found from
HE~1327$-$2326, however, strongly contradicts the WMAP value
($\log\epsilon ({\rm Li})=2.6$) from the baryon-to-photon ratio
(Spergel et al. 2007). This may indicates that the star formed from extremely
Li-poor material. No neutron-capture element is detected in
HE~0107$-$5240 or HE~0557$-$4840, whereas, unexpectedly, Sr is
observed in HE~1327$-$2326. Massive rotating stars may be responsible
for this neutron-capture element (Chiappini et al. 2011).

How can all those signatures be understood in terms of early chemical
enrichment?  HE~0107$-$5240 and HE~1327$-$2326 immediately became
benchmark objects to constrain various theoretical studies of the
early Universe, such as the formation of the first stars (e.g.,
Yoshida et al. 2006), the chemical evolution of the early ISM (e.g.,
Karlsson \& {Gustafsson 2005) or calculations of Pop\,III SN
yields. Their highly individual abundance patterns have been
successfully reproduced by several different SNe scenarios. This makes
HE~0107$-$5240 and HE~1327$-$2326 early, extreme Pop\,II stars that
possibly display the ``fingerprint'' of just one Pop\,III
SN. Umeda \& Nomoto (2003) first matched the yields of a faint
25\,M$_{\odot}$ SN that underwent a mixing and fallback process to the
observed abundances of HE~0107$-$5240.  To achieve a simultaneous
enrichment of a lot of C and only little Fe, large parts of the
Fe-rich SN ejecta have to fall back onto the newly created black
hole. Using yields from a SN with similar explosion energy and mass
cut, Iwamoto et al. 2005 then reproduced the abundance pattern of
HE~1327$-$2326 also. Trying to fit the observed stellar abundances,
Heger \& Woosley (2010) are employing an entire grid of Pop\,III SN
yields to search for the best match to the data. A similar progenitor
mass range as the Umeda \& Nomoto (2003) 25\,M$_{\odot}$ was found
to be the best match to have provided the elemental abundances to the
ISM from which these Pop\,II stars formed.

Limongi et al. (2003) were able to reproduce the abundances of
HE~0107$-$5240 through pollution of the birth cloud by at least two
SNe. Suda et al. (2004) proposed that the abundances of HE~0107$-$5240 would
originate from a mass transfer of CNO elements from a postulated
companion, and from accretion of heavy elements from the ISM. However,
neither HE~0107$-$5240 nor HE~1327$-$2326 show radial velocity
variations that would indicate binarity. Meynet et al. (2005) explored
the influence of stellar rotation on elemental yields of
60\,M$_{\odot}$ near-zero-metallicity SNe.  The stellar mass loss rate
of rotating massive Pop\,III stars qualitatively reproduces the CNO
abundances observed in HE~1327$-$2326 and other carbon-rich metal-poor
stars.

More generally, the observed abundances of the most metal-poor stars
with typical halo signatures in the range have successfully been
reproduced with Pop\,III SN yields. Tominaga et al. (2007b) model the
averaged abundance pattern of four non-carbon-enriched stars with
$-4.2<\mbox{[Fe/H]}<-3.5$ with the elemental yields of a massive,
energetic ($\sim30-50$\,M$_{\odot}$) Pop\,III hypernova. The
abundances can also be fitted with integrated yields of Pop\,III SNe
(Heger \& Woosley 2010). Special types of SNe or unusual
nucleosynthesis yields can then be considered for stars with
chemically peculiar abundance patterns. It is, however, often
difficult to completely explain the entire abundance pattern of a
given star. Additional metal-poor stars as well as a better
understanding of nucleosynthesis and the explosion mechanism and the
impact of the initial conditions on SNe yields are required to arrive
at a more solid picture of the details of early SNe nucleosynthesis.

%%%%%%%%%%%%%%%%%%%%%%%%%%%%%%%%%%%%%%%%%%%%%%
\section{Dwarf Archaeology}

Simulations of the hierarchical assembly of galaxies within the cold
dark matter (CDM) framework (Diemand et al. 2007; Springel 2005) show
that the Milky Way halo was successively built up from small dark
matter substructures, often referred to as galactic building blocks,
as long ago suggested by Searle \& Zinn (1978).
Figure~\ref{fig:allhalos} shows the substructure around the six
simulated, high-resolution ``Aquarius'' halos at $z=0$ (Springel et
al. 2008; Lunnan et al. 2011). The satellites around the main halos
are clearly visible -- these are smaller halos that survived the
violent accretion process until today.\footnote{In this simulation,
prescriptions for the population of dark halos with luminous matter
have already been applied. See Lunnan et al. (2011) for further
details and references therein.} They can be regarded as the
counterparts to today's satellite population of the Milky Way.

However, these simulations generally only include dark matter, and it
remains unclear to what extent small dark halos contain luminous
matter in the form of stars and gas. This question is particularly
important with respect to the so-called ``missing-satellite'' problem
which reflects the mismatch of the number of observed dwarf galaxies
surrounding the Milky with the predicted number of substructures for a
Milky Way-like halo. Studying the onset of star formation and
associated chemical evolution in dwarf galaxies thus provides some of
the currently missing information for our understanding of how the
observed properties of small satellites relate to the (dark matter)
substructures that build up larger galaxies. Thus, the study of the
entire stellar population of a dwarf galaxy for the purpose of
inferring details about the nature and origin of the first galaxies
and early galaxy assembly is termed ``dwarf archaeology''.

%----------------------------------------------
\begin{figure*}[!t]
\center{
\includegraphics[width=1.\textwidth]{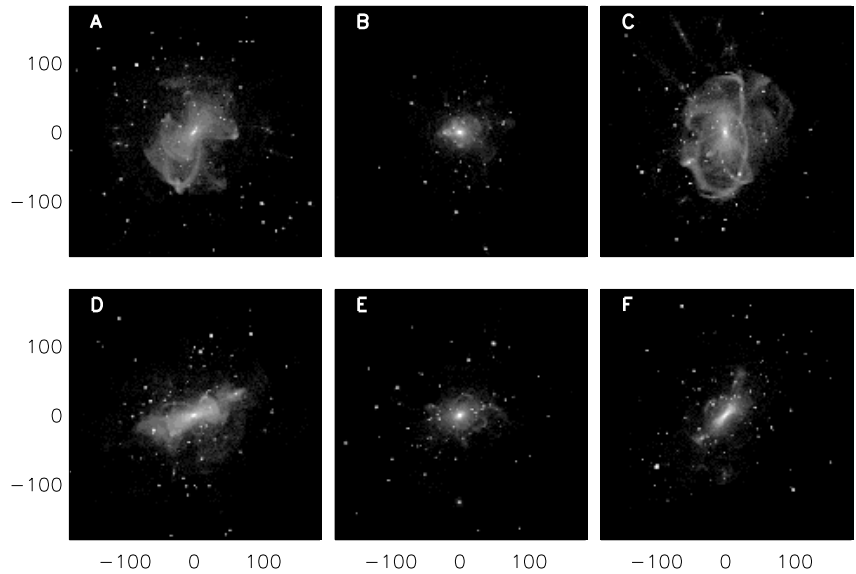}
}
\caption{\label{fig:allhalos}
Projected spatial distribution of surviving substructure, as well as
accreted material that forms a stellar halo, in the six Aquarius
simulations. The scale is in kpc. Figure taken from Lunnan et al. (2011).}
\end{figure*}
%----------------------------------------------

\subsection{Chemical Evolution of Dwarf Galaxies} 

The connection between the surviving dwarfs and those that dissolved
to form the halo can partially be addressed by examining in detail the
stellar chemical compositions of present-day dwarf
galaxies. Establishing detailed chemical histories of these systems
can provide constraints on their dominant chemical enrichment events
and timescales. From such information, conclusions about the formation
process of the Milky Way can be derive. Specifically, detailed
knowledge of the most metal-poor stars in a given system allow insight
into its earliest phases of star formation before the effects of
internal chemical evolution were imprinted in stars born later with
higher metallicity. Despite the fact that dwarf galaxies are regarded
as ``simple'' systems, many of them show extended star formation
histories with similar, albeit slower, chemical evolution to the Milky
Way (for further details on this topic the reader is referred to the
recent reviews by Tolstoy et al. 2009 and
Koch 2009). Altogether, they are old, gas-less systems, and
a tight correlation between the average metallicity and systemic
luminosity.

%\cite{kirby08} shows the faintest galaxies in the universe ($L_{\rm
%tot} \leq 10^{5}\,L_{\odot}$) to be the most metal-poor systems
%($\mbox{[Fe/H]}_{ave}\sim-2.6$), comparable only to the most
%metal-poor stars in the Milky Way halo.

Kirby et al. (2008) extended the metallicity-luminosity relationship to the
ultra-faint ($L<10^{5}\,L_{\odot}$; discovered in the Sloan Digital
Sky Survey; SDSS) dwarfs. They also showed that the lowest luminosity
dwarfs have very low average metallicities (down to
$<\mbox{[Fe/H]}>\sim-2.6$), with several systems having values lower
than those of the most metal-poor globular clusters. Many individual
stars even have $\mbox{[Fe/H]} < -3$, while interestingly, no stars
with $\mbox{[Fe/H]}>-1.0$ are found. The combined MDF of all the
ultra-faint dwarfs extends down to $\mbox{[Fe/H]}\sim-3.8$, and the
shape appears similar to that of the Milky Way halo (for the
low-metallicity tail, that is), although perhaps with a slightly
larger relative fraction towards the most metal-poor stars.  Thus, it
is not surprising that the ultra-faint dwarf galaxies contain a
relatively many of extremely metal-poor stars, with large internal
[Fe/H] abundance spreads of up to 3\,dex. These spreads were first
found in lower spectral resolution data (e.g., Simon \& Geha 2007;
Norris et al. 2008) and later confirmed with high-resolution follow-up
of individual stars. At the same time, it became apparent that
$\sim30\%$ of the known stars with $\mbox{[Fe/H]}<-3.5$ are now found
in dwarf galaxies. Segue\,1, the most metal-poor system, contains
$\sim15\%$ alone (Frebel et al. in prep.).

Such high-resolution spectra are needed for a full chemical abundance
analysis. Given the faint magnitude of stars in dwarf galaxies, these
data are difficult to acquire, but good progress has been made over
the past few years. In fact, most stars with $V<19.2$ available in the
ultra-faint dwarfs have now been observed this way. More ultra-faint
dwarfs are expected to be found soon in surveys such as Pan-Starrs,
Skymapper and LSST, thus providing new targets that can again be
observed with high-resolution on current 6-10\,m telescopes.

Generally, these kinds of data enable measurements of chemical
abundances and upper limits of 10-15 elements for a given star.  For
example, the three brightest stars in each of Ursa Major\,II (UMa\,II)
and Coma Berenices (ComBer) and two stars in Hercules are the first
stars in the ultra-faint dwarf galaxies to have been observed with
high-resolution spectroscopy. Two of them (in UMa\,II) are also the
first known extremely metal-poor stars which are not members of the
halo field population. Details on the observations and analysis
techniques are given in Frebel et al. (2010a) and Koch et
al. (2008). For the UMa\,II and ComBer stars, chemical abundances and
upper limits of up to 26 elements were determined for each star. The
abundance results demonstrate that the evolution of the elements in
the ultra-faint dwarfs is very similar to that of the Milky Way, and
likely dominated by massive stars. The $\alpha$-elements in these two
ultra-faint dwarf stars are overabundant, showing the halo-typical
core-collapse SNe signature (see Figure~\ref{alphas}).

The neutron-capture abundances are extremely low in the ultra-faint
dwarf stars (see green (large shaded) circles in Figure
~\ref{ncap}). The observed Sr and Ba values are up to two orders of
magnitude below the abundances found in typical MW halo stars with
similar Fe content. However, a very large spread (up to 3\,dex) in
these elements is found among halo stars themselves. The large
depletions in the dwarf galaxy stars are thus not inconsistent with
the halo data since similarly low values are found in numerous
objects. Interestingly though, the low neutron-capture abundances may
represent a typical signature of stars in dwarf galaxies. Similarly
low values are also found in Hercules (Koch et al. 2008) and Draco
(Fulbright et al. 2004) despite their sometimes relatively high Fe
values of $\mbox{[Fe/H]}\sim-2.0$ (in Hercules).

\begin{figure*} [!t]
\center{
\includegraphics[width=11.7cm,clip=true,bbllx=45,bblly=343,bburx=443,bbury=657]{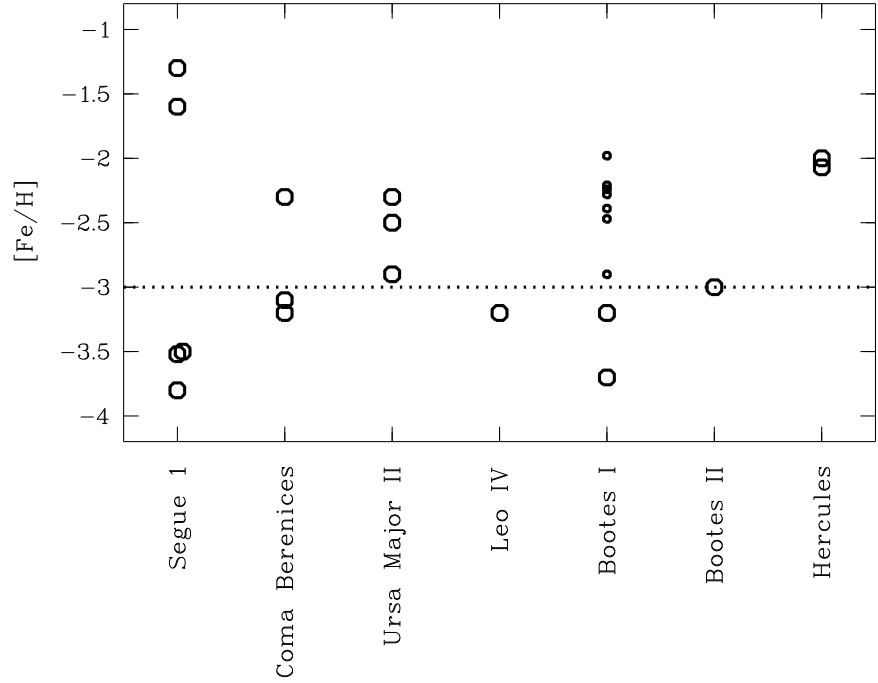}}
\caption{\label{fig:status} Current status of available
high-resolution spectroscopy of stars in the ultra-faint dwarf
galaxies. Many stars have $\mbox{[Fe/H]}<-3.0$ Large circles indicate extensive abundances studies of many
elements, whereas smaller circles refer to studies of only a few
elements. Data from 
Koch et al. (2008),
Feltzing et al. (2009),
Frebel et al. (2010a),
Simon et al. (2010),
Norris (2010b),
Norris et al. (2010a) and
Frebel et al. (2011, in prep.)
}
\end{figure*}

\begin{figure*} [!t]
\center{
\includegraphics[width=11.7cm,clip=true,bbllx=27,bblly=265,bburx=478,bbury=732]{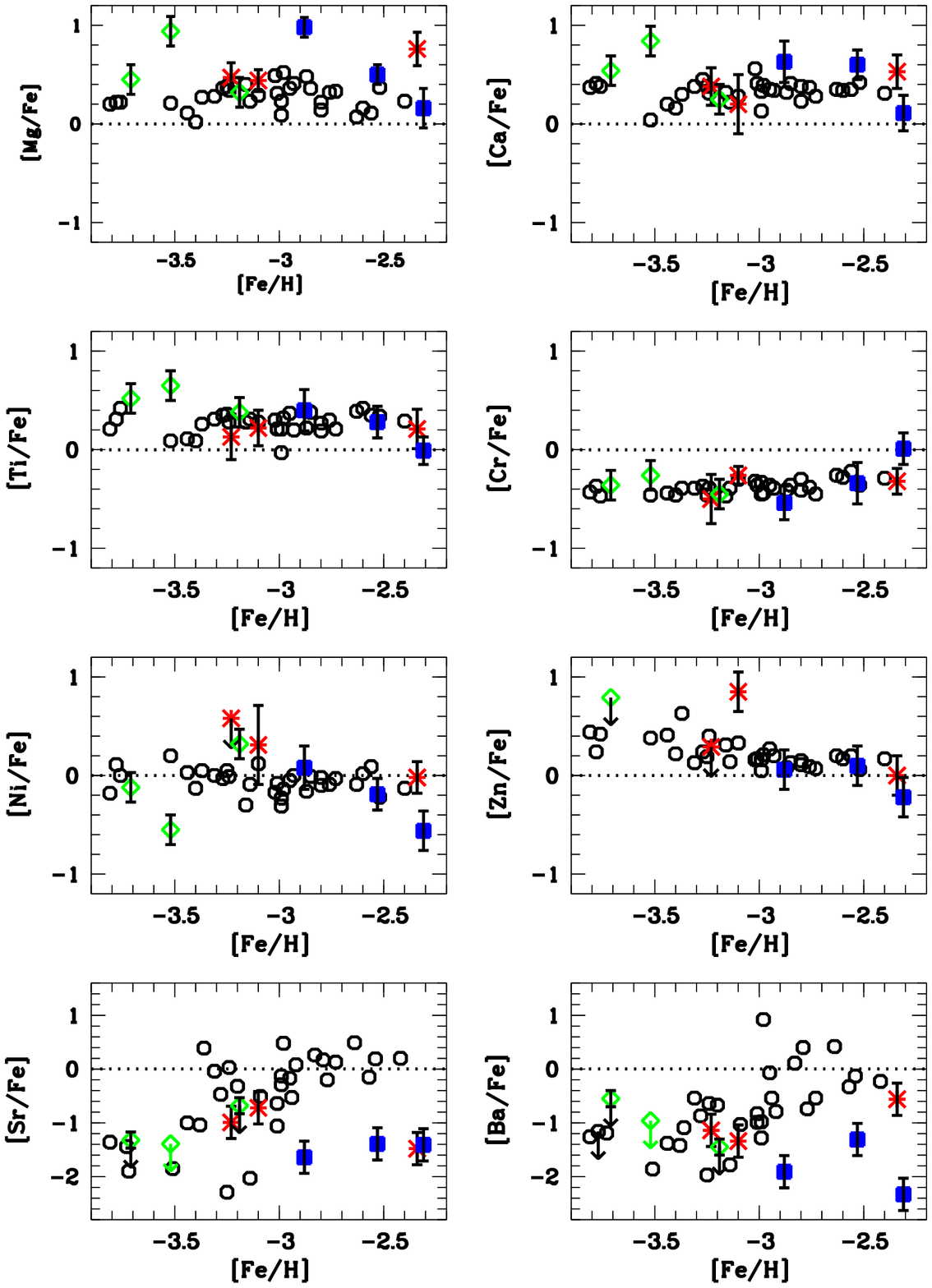}}
\caption{\label{fig:abund} Chemical abundance comparison of Galactic
halo stars (open circles, from Cayrel et al. 2004) and stars in the
ultra-faint dwarfs (asterisks: Ursa Major\,II, filled squares: Coma
Berenices, diamonds: Segue\,1, Bootes\,I and Leo\,IV). The abundances
generally agree very well, suggesting a similar enrichment history of
the gas from which all these stars formed. Based on Figure~20 from
Frebel \& Norris (2011).}
\end{figure*}

By applying improved search techniques also to the more luminous dwarf
galaxies, the first extremely metal-poor star in a classical dwarfs
Sculptor was recently discovered (in a sample of 380 stars;
Kirby et al. 2009). The metallicity of $\mbox{[Fe/H]} \sim -3.8$ was
confirmed from a high-resolution spectrum taken with Magellan/MIKE
(Frebel et al. 2010b). Only nine stars in the halo have even lower Fe abundances
than this object. Shortly afterwards, Tafelmeyer et al. (2010) presented
additional, similarly metal-poor stars in Sculptor, and also Fornax.
These discoveries show that such low-metallicity stars are indeed
present in the more luminous dwarfs (see also Starkenburg et al. 2010),
contrary to earlier claims.

The new stars underpin that metallicity spreads of $\sim3$\,dex or
more are present in many classical dwarfs. The chemical abundances
obtained from the high-resolution spectrum reveal a similar picture to
what has been found in the ultra-faint dwarf stars. For example, the
Sculptor star at $\mbox{[Fe/H]}\sim-3.8$, also has a remarkably
similar chemical make-up compared to that of the Milky Way halo at the
lowest metallicities.  This is in contrast to what is found at higher
metallicities in these brighter dwarfs. As can be seen in
Figure~\ref{alphas} (red (small shaded) circles), those stars have
lower stellar $\alpha$-abundance ratios than comparable halo stars,
indicating different enrichment mechanisms and timescales in these
gas-poor systems (e.g., Shetrone et al. 2003; Geisler et al. 2005).

\subsection{Individual Pop\,III Enrichments?}\label{ind}

The large number of extremely metal-poor stars in the faintest dwarf
galaxies is very surprising given the very limited total stellar
material in these systems with $L_{tot} \lesssim 10^{5} L_{\odot}$.
This represents an opportunity to study the environment of a
low-metallicity galaxy which should have hosted only a few
SNe. Consequently, signs of individual, stochastic enrichment events
may be preserved in the metal-poor stars.

Enrichment by a single 13\,M$_{\odot}$ has been suggested for the
Hercules dwarf ($L \sim 10^{5} L_{\odot}$), based on the chemical
abundances of two stars with $\mbox{[Fe/H]}\sim-2.0$ and unusually low
levels of Ba (Koch et al. 2008). With its low luminosity of ($L_{V} =
14,000 L_{\odot}$) and low metallicity, Simon et al. (2010) determined
that Leo\,IV contains only 0.042 M$_{\odot}$ of Fe (assuming a stellar
mass-to-light ratio of 1 M$_{\odot}$/L$_{\odot}$).  Canonical SN
nucleosynthesis yield calculations predict a maximum Fe yield of
$\sim0.1M_{\odot}$ (e.g. Heger \& Woosley 2010). The total Fe amount
in Leo\,IV, and by extension that of all the other observed elements,
could thus stem from a single star that exploded soon after the system
formed. The star forming gas cloud at that time must have been only
$\sim 40,000 M_{\odot}$. Given that the chemical abundances of the
brightest star that is accessible to high-resolution spectroscopic
studies, resemble those of metal-poor halo stars, it was suggested
that indeed a single Pop\,III star was responsible for Leo\,IV's
enrichment. Various SN nucleosynthesis models invoking different
progenitor masses and explosion energies (e.g., Tominaga et al. 2007b;
Heger \& Woosley 2010) have shown that the typical halo abundance
signature can be explained this way.

Alternatively, given the shallow potential of these kinds of systems,
several SNe could have contributed to the enrichment, but part of the
enriched gas was soon lost (e.g., through stellar winds or SN
explosions). Future observations of stars in ultra-faint dwarfs will
provide more detailed insight. Nevertheless, the currently available
data already suggest that these systems played an important role for
our understanding of the first stars and chemical enrichment events
that shaped the nature of these relatively small, early systems.

\subsection{Are the faintest dwarf satellites surviving first galaxies?}

Given that the ultra-faint dwarf galaxies are the most metal-poor
objects we know of today, they are ideal probes of the physical,
chemical, and dynamical processes at work in the early universe.
Since these systems have much simpler star formation histories than
the halo of the Milky Way, their stellar populations should preserve
the fossil record of the first supernova (SN) explosions in their
long-lived, low-mass stars (see also Section~\ref{ind}).  Hence, the
metal-poor stars in the ultra-faint dwarf galaxies should be used to
empirically constrain the formation process of the first galaxies, and
early galaxy assembly more generally.

Guided by recent hydrodynamical cosmological ``ab-initio''simulations
(Greif et al. 2008; Greif et al. 2010) of first galaxy formation,
Frebel \& Bromm (2011) developed a set of stellar abundance signatures
that characterizes the nucleosynthetic history of such an early system
if it was observed in the present-day universe.  In particular, the
simulations suggest that a first galaxy can be regarded a chemical
``one-shot'' event, where only one (long-lived) stellar generation
forms after the first, Population\,III, SN explosions. The system
would thus be dominated by an [$\alpha$/Fe] enrichment due to
enrichment by massive stars as seen in the halo at low
metallicity. With no stars present displaying an erstwhile enrichment
by AGB stars of SN\,Ia, the $\alpha$-enhancement would also be present
in stars with higher metallicities ($\mbox{[Fe/H]}>-1.5$).  These
criteria thus constrain the strength of negative feedback effects
inside the first galaxies.

A comparison of the abundances of about a dozen stars in the
ultra-faint dwarfs with this one-shot criterion indicates that some of
these faintest satellites could be surviving first galaxies. Several
systems (Ursa Major\,II, and also Coma Berenices, Bootes\,I, Leo\,IV,
Segue\,1) largely fulfill the requirements (most notably the high
$\alpha$/Fe ratios), indicating that their high-redshift predecessors
did experience strong feedback effects that shut off star formation
soon after the formation of the system. More observations are needed
to firm up these initial conclusions, and also additional simulations
of early metal mixing, turbulence and the extend of feedback in early,
assembling galaxies. This will provide clues to the connection of the
first galaxies, the surviving, metal-poor dwarf galaxies, and the
building blocks of the Milky Way.

\section{Near-field cosmology}

A long-standing problem, originally noted by e.g. Moore et al. (1999) and
Klypin et al. (1999), is that the observed number of Milky Way satellites
appears to be significantly lower than the number of dark matter
substructures expected based on the CDM theory (the so-called ``missing
satellite problem'').  This calls into question the validity of the CDM
picture on the scales of individual galaxies.  Many ideas have been
proposed to reconcile the theory with the observations, including the
possibility that the dark matter may instead be dynamically ``warm'',
rather than cold (achieved by making the individual dark matter
particles less massive).  Another class of models proposes instead
that effects related to feedback from baryonic processes or
heating by cosmic radiation fields may inhibit star formation in dark
matter halos of sufficiently low mass, rendering them invisible.
Either way, a definitive solution to the missing satellite problem
will inform our theories of the nature of the dark matter and the
assembly of galaxies.

Progress has been made through detailed studies of different types of
dwarf galaxies (ultra-faint, classical dwarf spheroidal, more massive
dwarfs such as the Magellanic Clouds) which orbit the Milky
Way. Extensive photometric and spectroscopic data of these satellites,
paired with the discovery of stellar streams in the Galactic halo
arising from past and even ongoing mergers of the massive host with
smaller galaxies, have revealed much about the complex nature of the
life and times of dwarf galaxies and their role in shaping their
parent galaxy. Figure~\ref{fig:fos} shows a number of streams that are
currently present on the stellar halo of the Milky Way due to past
accretion events.

Nevertheless it is surprising that the Milky Way overall has so few
satellite galaxies when its sister galaxy Andromeda appears to have a
significantly larger population signaling a potentially very different
assembly history. By learning about the Galaxy and its assembly, as
well as pushing for large-scale simulations to address cosmic variance
and where the Milky Way really fits within the zoo of galaxies, we
will soon be able to quantify the host of observations that suggest
the unique nature of our Galaxy.

\begin{figure} [!t]
\includegraphics[width=1.\textwidth]{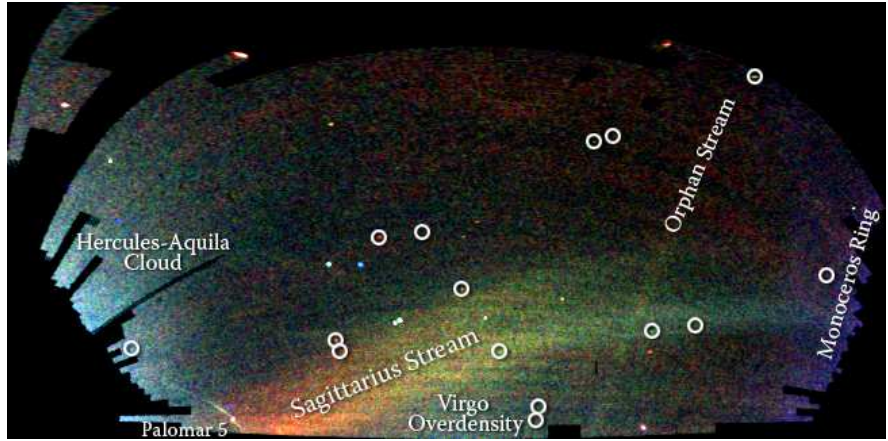}
\caption{\label{fig:fos} Field of Streams:  
A map of stars in the outer regions of the Milky Way Galaxy, shown
in a Mercator-like projection. The color indicates the distance of the
stars, while the intensity indicates the density of stars on the
sky. Structures visible in this map include streams of stars torn from
the Sagittarius dwarf galaxy, a smaller ``orphan'' stream crossing the
Sagittarius stream, the ``Monoceros Ring'' that encircles the Milky Way
disk, trails of stars being stripped from the globular cluster Palomar
5, and excesses of stars found towards the constellations Virgo and
Hercules. Circles enclose new Milky Way companions discovered by the
SDSS; two of these are faint globular star clusters, while the others
are faint dwarf galaxies. 
Image credit: V. Belokurov and the Sloan Digital Sky Survey.}
\end{figure}

\subsection{Did the stellar halo form from accreted dwarf galaxies?}

The overarching questions of near-field cosmology concern the nature
of the ``building blocks'' of large galaxies, and to what extent dwarf
galaxies play a role in the assembly of old stellar halos. This comes
at a time when in particular the population of Milky Way ultra-faint
dwarf galaxies have been shown to be extremely metal-deficient systems
that host $\sim30\%$ of the known most metal-poor stars.  Moreover,
they extend the metallicity-luminosity relationship of the classical
dwarfs down to $L_{\rm tot} \sim 10^{3}\,L_{\odot}$ (see
Kirby et al. 2008 for more details). Future observations will reveal how
far this relationship can be extended.

If the old, metal-poor halo was indeed assembled from dwarf galaxies,
the metallicities of stars in dwarf galaxies must reach values as low
as (or lower) what is currently found in the Galactic halo
population. Moreover, the abundance ratio of those low-metallicity
stars must be roughly equal to those of equivalent stars in the halo.

Earlier works missed finding extremely metal-poor stars in the
classical dwarfs, which posed a significantly problem to the idea of
an accreted halo through systems similar to the surviving dwarfs.
However, it has now been shown that this claim stems merely from
biases in earlier search techniques (Cohen et al. 2009; Kirby et
al. 2008a; Starkenburg et al. 2010; Kirby et al. 2010). With improved
methods for identifying the lowest-metallicity objects, extremely
metal-poor stars have already been identified in several dwarf
galaxies (Kirby et al. 2008; Geha et al. 2009; Frebel et al. 2010b;
Tafelmeyer et al. 2010). The higher-metallicity stars in the classical
dwarfs all have abundance ratios different from comparable halo
stars. Most strikingly, the $\alpha$-element abundances are not
enhanced, which must be due to the onset on SN\,Ia enrichment at a
time when the overall chemical evolution of these small systems was
less evolved than the halo. It also reflects that any of these stars
are the product of the galaxy's own chemical evolution that has
nothing in common with that of the Milky Way. However, there is
increasing evidence that a transition to more halo-typical abundance
ratios may take place around a metallicity of $\mbox{[Fe/H]}= -3.0$
(Cohen et al. 2009; Aoki et al. 2009) in these dwarfs. This means that
if the halo formed from accreted satellites, any systems like the
classical dwarfs must have been accreted at a relatively early time
when the chemical evolution had not much progressed. Given that the
dwarfs we observe today were not accreted, they kept forming stars and
continued chemical evolution until they lost all their gas.

Due to their simple nature, the ultra-faint systems are expected to
retain signatures of the earliest stages of chemical enrichment in
their stellar populations. If the halo was assembled from early
analogs of the surviving dwarfs, most of the metal-poor halo stars
should have been deposited there during late-time accretion events
(Frebel et al. 2010a; Simon et al. 2010; Font et al. 2006). Indeed the
chemical abundances of individual stars in the faintest galaxies
suggest a close connection to equivalent, extremely metal-poor halo
stars in the Galaxy.

The recent high-resolution studies (Feltzing et al. 2009; Frebel et
al. 2010a; Norris et al. 2010a,2010b; Simon et al. 2010) provide
evidence that the abundance patterns of light elements ($Z < 30$) in
these dwarfs are remarkably similar to the Milky Way halo, even for
stars with $\mbox{[Fe/H]}> -3.0$. this is illustrated in
Figure~\ref{fig:abund} which shows a very detailed comparison between
halo and ultra-faint dwarf galaxy stars. The similarity between the
data sets is clearly seen. However, given the limited number of stars
it is still unclear up to what metallicity the halo-typical abundances
are found in these systems.  There are indications that the chemical
evolution in the ultra-faint dwarfs may have been very inhomogeneous,
and also, that a number of stars show strong carbon-enhancement
(Norris et al. 2010a). Taken altogether, these features are found
among the lowest metallicity halo stars, making a plausible case for
an accretion-built halo, at least for stars with $\mbox{[Fe/H]}<
-2.5$.

These observational results about the halo assembly are broadly
consistent with the predictions of CDM cosmological models
(e.g. Robertson et al. 2005; Johnston et al. 2008). While the majority of
the mass that formed the inner part of the stellar halo (at
$\mbox{[Fe/H]}\sim-1.2$ to $-1.6$) was provided at early times by much
larger systems such as the Magellanic Clouds, the outer halo assembled
at later times. In fact, as shown in Figure~9, it is still assembling
today, with ongoing accretion event leading to a variety of streams
and substructure. The similarity of the chemical abundances,
suggesting the same chemical enrichment history in these stars prior
to their formation, make it plausible to assume that the ultra-faint
dwarf galaxies contributed (some) individual metal-poor stars to the
Galaxy. These stars are now found primarily in the outer Galactic
halo.

However, these systems may not have been sufficiently numerous to
account for the entire metal-poor end of the Fe metallicity
distribution of the Milky Way halo. Since the classical dSphs have
more stellar mass and have been shown to also contain at least some of
the most metal-poor stars (e.g., Kirby et al. 2009; Frebel et
al. 2010b), they could have been a major source of the
lowest-metallicity halo stars.  Additional observations of more
extremely metal-poor stars in the various dwarf galaxies are highly
desirable in the quest to determine individual MDFs for each of these
galaxies, and how those would compare with each other and that of the
Milky Way.

\subsection{Towards constraining the reionization history of the Milky
Way}

Cosmological simulations of the growth of structure have shown that
galaxy assembly proceeds hierarchically.  The results of these studies
have indicated a mismatch between the predicted number of low-mass,
dark matter substructures and the actual observed number of satellite
galaxies around the Milky Way; the ``missing satellite problem''.  The
underlying cause of this discrepancy can be investigated with
cosmological simulations. By examining the impact of various physical
processes on the evolution of faint galaxies, the nature of the
surviving luminous satellites and their stellar content can be
understood in more detail. An important question is this regard is
which effects critically determine the number of small satellites
hosting luminous matter throughout the build-up of a large galaxy, and
whether there would be enough of them to form a metal-poor stellar
halo of their host, similar to what is found for the Milky Way.

One explanation for the small number of faint satellites is that as
the universe reionizes, the increased temperature of the IGM prevents
the smallest halos from accreting and cooling gas, preventing further
star formation. In this scenario, the nature and abundance of the
faintest dwarf galaxies can in principle provide a constraint on the
reionization history of the Milky Way (Munoz et al 2009, Busha et
al. 2010, Lunnan et al. 2011).
%% Assuming that reionization of the universe was patchy and extended in
%% redshift (Zahn et al. 2007; McQuinn et al. 2007), the number of
%% smallest, resolved satellites ($10^{6}$\,M$_{\odot}$ halos) that
%% survive until z=0 can reduced relative to models in which reionization
%% is considered to be homogeneous and instantaneous (Lunnan et
%% al. 2011). This is about a 10\% effect. 
This is illustrated in Figure~\ref{fig:lum_fct}, which shows observed
and predicted luminosity functions of dwarf galaxies for the six Milky
Way-like halos from the Aquarius simulation, combined with four
different reionization models. The dotted line shows the prediction
with no reionization effects, clearly illustrating the ``missing
satellite'' problem. This is the case for all six halos. The general
halo-to-halo scatter (factor of 2-3; see also
Figure~\ref{fig:allhalos}), however, suggests that the missing
satellite problem is in part due to cosmic variance. Moreover, taking
the effects of reionization into account has a pronounced impact of
just the faintest halos. While larger halos (equivalents to today's
brighter dwarf galaxies) are relatively insensitive to a
non-instantaneous reionization prescription, the number of surviving
small galaxies changes significantly with the reionization model for
different mean reionization redshift and Thompson optical depths still
in agreement with the WMAP value (Komatsu et al. 2011).

Given these initial results, it appears promising to use the brighter
end of the luminosity function to constrain the halo-to-halo scatter,
while the faint end will enable to discriminate reionization
histories. However, a solid understanding of how ``normal'' the Milky
Way is and where falls within these halo-to-halo variations and will
be required. Future simulations may be able to quantify the nature and
being of the Galaxy with respect to the majority of other large
galaxies. Knowing the substructure abundance of galaxy halos is
critical for interpreting observations of the satellite populations of
all large galaxies, including the Milky Way and Andromeda. Moreover,
estimates would become possible of the stellar contribution to the
halo as a function of halo mass and the associated merger and
accretion history. 

Altogether, the opportunity to constrain the reionization history of
the Milky Way through a careful analyses of the faintest satellite
population will help understanding the nature of these systems
themselves. Knowing the fraction of surviving small halos with truly
old stars that formed before reionization is of critical importance so
it can be established that today's metal-poor stars indeed trace the
earliest times and enrichment events. Using the cosmological
simulations together with prescriptions for luminous matter,
feedback processes and chemical evolution (e.g., Cooper et al. 2010) will
shed more light on the cosmological origin of the most metal-poor
stars. Stellar and dwarf archaeology meet near-field cosmology for
exactly these kinds of questions, and only the combination of
high-quality observations and powerful cosmological simulations will
enable the progress that is required to understand the early star and
galaxy formation and the evolution of our own Milky Way.

%----------------------------------------------
\begin{figure*} [!t]
\center{
\includegraphics[width=1.04\textwidth]{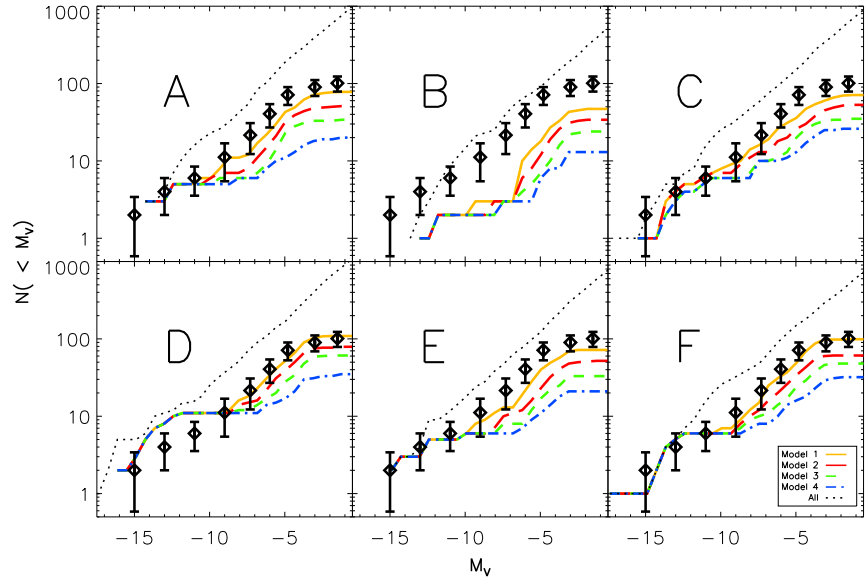}}
\caption{\label{fig:lum_fct} Observed luminosity function of dwarf
  galaxies (black diamonds) in comparison with models employing four
  different physically motivated reionization histories for the six
  Aquarius halos. Note the scatter in the numbers of simulated
  satellites among the halos (factor of 2-3) as well as similar
  variations of the faint end caused by the different reionization
  histories. The dotted lines depict satellite population unaffected
  by reionization. From Lunnan et al. (2011).}
\end{figure*}

\section{Outlook}

Old metal-poor stars have long been used to learn about the conditions
of the early Universe.  This includes the origin and evolution of the
chemical elements, the relevant nucleosynthesis processes and sites
and the overall chemical and dynamical history of the Galaxy. By
extension, metal-poor stars provide constraints on the nature of the
first stars and their initial mass function, the chemical yields of
first/early SNe, as well as early star and galaxy formation processes
including the formation of the galactic halo.

Supported by these large-scale survey efforts, the field of near-field
cosmology has been catapulted into an ``excited state'' because it
constrains many astrophysical problems related to galaxy formation. It
is thus very timely to extensively pursue research into the oldest
stars and stellar populations in the dwarf galaxies. They not only
enable us to study the history of galaxies with different masses and
luminosities, but also facilitate an in-depth study of what role
small(er) galaxies may have played in the build-up of the stellar
halo.

\subsection{Observational Challenges}
 Finding more of the most metal-poor stars (e.g., stars with
$\mbox{[Fe/H]}<-5.0$) is of great importance for addressing the topics
described in this chapter in more detail. However, as outlined these
stars are extremely rare (Sch\"orck et al. 2009) and difficult to find. The
most promising way forward is to survey larger volumes further out in
the Galactic halo. But how feasible is it to identify stars with even
lower metallicities?  Frebel et al. (2009) calculated the minimum observable Fe
and Mg abundances in the Galaxy by combining the critical metallicity
of $\mbox{[C/H]}_{min}=-3.5$ (the criterion for the formation of the
first low-mass stars by Bromm \& Loeb (2003) with the maximum
carbon-to-iron ratio found in any metal-poor star. The resulting
minimum Fe value is $\mbox{[Fe/H]}_{min}=-7.3$. Analogously, the
minimum Mg value is $\mbox{[Mg/H]}_{min}=-5.5$. If
$\mbox{[C/H]}_{min}$ was lower, e.g., $\mbox{[C/H]}_{min}=-4.5$, as
suggested by recent dust cooling computations, the minimum observable
Fe and Mg abundances would accordingly be lower. Spectrum synthesis
calculations suggest these low abundance levels are indeed measurable
from each of the strongest Fe and Mg lines in suitably cool
metal-deficient giants.

Future surveys such as  Skymapper (Keller et al. 2007) and LAMOST will provide
an abundance of new metal-poor candidates as well as new, faint dwarf
galaxies. By accessing such stars in the outer Galactic halo and dwarf
galaxies we will be able to gain a more complete census of the
chemical and dynamical history of our own Galaxy. Since the lowest
metallicity stars are expected to be in the outer halo (e.g.,
Carollo et al. 2011), their corresponding kinematic properties may prevent
them from accreting too much enriched material from the ISM during
their lives so that their surface composition would not be altered
(i.e., increased; Frebel et al. 2009). Hence, selecting for the most
metal-poor candidates will increasingly rely on our ability to combine
chemical abundances with kinematic information. Future
missions such as GAIA will provide accurate proper motions for many
object that currently have no kinematic information available, including for
most of the currently known metal-poor giants.

However, many, if not most, of these future metal-poor candidates will
be too faint to be followed up with the high-resolution spectroscopy
necessary for detailed abundance analyses. This is already an issue
for many current candidates leaving the outer halo a so far largely
unexplored territory: The limit for high-resolution work is
$V\sim19$\,mag, and one night's observing with 6-10\,m telescopes is
required for the minimum useful signal-to-noise ($S/N$) ratio of such
a spectrum. With the light-collecting power of the next generation of
optical telescopes, such as the Giant Magellan Telescope, the thirty
Meter Telescope or the European ELT, and if they are equipped with
high-resolution spectrographs, it would be possible to not only reach
out into the outer halo in search of the most metal-poor stars, but
also provide spectra with very high-$S/N$ ratio of somewhat brighter
stars. For example, the so-called r-process enhanced stars which
provide crucial empirical constraints on the nature of this
nucleosynthesis process require exquisite data quality e.g. for
uranium and lead measurements that are currently only possible for the
very brightest stars (e.g., Frebel et al. 2007).

It appears that the hunt for the metal-deficient stars in dwarf
galaxies may have just begun since these dwarfs host nearly a third of
the known low-metallicity stars.. The detailed abundance patterns of
the stars in UMa\,II, ComBer, Leo\,IV, etc.  are strikingly
similar to that of the Milky Way stellar halo, thus renewing the
support for dwarf galaxies as the building blocks of the halo. Future
discoveries of additional faint dwarf galaxies will enable the
identification of many more metal-poor stars in new, low-luminosity
systems.  But also the brighter dSphs have to be revisited for their
metal-poor content (Kirby et al. 2009).  More stars at the lowest
metallicities are clearly desired to better quantify the emerging
chemical signatures and to solidify our understanding of the early
Galaxy assembly process. Together with advances in the theoretical
understanding of early star and galaxy formation and SNe yields, a
more complete picture of the evolution of the Milky Way Galaxy and
other systems may soon be obtained. Only in this way can the
hierarchical merging paradigm for the formation of the Milky Way be
put on firm observational ground.

\subsection{Constraining the Theoretical Framework}

An important next step towards a full understanding of the infant
universe is to combine the phenomenological approach of collecting new
observational data with theoretical knowledge about the formation of the first
generations of stars and galaxies. This way, tools can be developed
that allow for cosmologically motivated interpretations of the
abundance patterns of metal-poor stars, in both the halo and dwarf
galaxies. The new field of dwarf archaeology promises a more
complete understanding of early enrichment events and the processes
that led to galaxy formation at the end of the cosmic dark ages. At the
same time, this will have profound implications for the search of the
major enrichment mechanisms in the early universe and the 
physical origin of these stars.

Understanding the Milky Way as a whole is thereby of crucial
importance. Most importantly, new results have raised the question of
how good the Milky Way of in representing a typical, large spiral
galaxy.  Our Galaxy is often used as a reference, especially when
comparing its general properties with those derived from simulations
of the formation of large galaxies. But recent works, both
observationally and theoretically, have shown that our Galaxy has at
least several unusual features. The existence of the long discovered
Magellanic Cloud satellites have recently garnered significant
attention (Boylan-Kolchin et al. 2010, Liu et al. 2011) in this
respect. Observational analyses using Sloan Digital Sky Survey data of
many other large spiral galaxies confirmed that galaxies like the
Milky Way are very unlikely to have two companions as bright as the
Magellanic Clouds. Indeed, less than 5\% of galaxies host two such
bright companions, and more than 80 percent host no such satellites at
all. Previously, Boylan-Kolchin et al. (2010) have examined this issue
using the Millennium simulation of Springel et al. (2005) and finding
that the Milky Way was unusual in hosting the Magellanic Clouds.

Furthermore, the evolutionary differences between the Milky and
Andromeda need to be to established. The Milky Way did not undergo
major mergers with other galaxies since nearly 11 billion years,
whereas Andromeda underwent many mergers in a recent past (a few
billion years). Only then, can be understood how the different
formation histories influence the extent of surviving substructure and
differences in properties such as the stellar mass, disk radius, and
metal-deficient halo between the sister two galaxies.
Assessing the degree of ``normality'' of the Milky Way will be vital for
understanding of whether or not the Magellanic Clouds were only
recently accreted by the Milky Way, as proposed by Besla et al. (2007)
based on new proper motion measurements.

A major step forward would be a detailed understanding of the missing
satellite problem.  This is of great interest in cosmology, both
observationally and theoretically, but also many associated fields
e.g, regarding galaxy formation or dwarf galaxy studies. Is the
overproduction of halo substructure at $z=0$ in DM simulations really
based on physical processes associated with the gas that lights up
dark halos, or is it merely an artifact of past simulations? It has
been shown that cosmic variance may play an important role, and
without properly quantifying this effect, perhaps no conclusions can be
drawn at this point in time. Nevertheless, the fact that the Milky Way
may not be an ordinary galaxy may partly explain the missing satellite
problem as all simulation results are always compared to the
observations associated with the Galaxy.

Progress can be made by carrying out simulations that aim at
incorporating near-field cosmological constraints to address, e.g.,
the underlying physical causes of the missing satellite problems
beyond cosmic variance. To learn about the effects that influence the
number of small subhalos over the course of the universe, studies are
now being carried out that aim at quantifying the impact of physically
motivated, patchy, reionization histories on the faintest halos (e.g.,
Lunnan et al. 2011). While these studies may partly resolve the missing
satellite problem, the overall halo-to-halo differences in the
populations is of a similar level, preventing strong, global
conclusions.  The way forward it to quantify the level and extend of
variations of the substructure around Milky Way galaxies. Only then a
number of specific details about the origin and evolution of subhalos
with different masses equivalent to those of a variety of observed
dwarf galaxies, perhaps analogous to  massive Magellanic Cloud-sized objects,
classical dwarf Spheroidal galaxies, and even fainter systems, can be
understood.

%Tied to this issue is the question of the role of the different types
%of faint and luminous satellite galaxies in the build-up of large
%galaxies like the Milky Way and Andromeda, and how observations of the
%numerous stellar streams can be interpreted within the hierarchical
%assembly process of the Galaxy.

In summary, the details of the many physical processes that govern the
evolution of a luminous halo at z=0 will need to be known to
conclusively address the formation of large galaxies within the
hierarchical assembly paradigm. They will need to be coupled to
simulation results, quantified, and compared to the
observations. Hopefully, in the not too distant future, there will be
unparalleled opportunities to study the assembly of galaxy halos in
close connected to the results of the latest observations of dwarf
galaxies, halo stars and stellar streams found in the Milky Way. For
example, [Fe/H] spreads in dwarf galaxies and well-established
abundance trends are important to constrain the chemical evolution in
dwarf galaxies. These quantities are crucial for the development of
prescriptions for the chemical enrichment throughout the hierarchical
merging process. Moreover, they constrain, e.g., metal mixing,
turbulence and feedback effects in early star forming halos, as
simulated in hydrodynamical ``ab-initio'' simulations (e.g.,
Greif et al. 2010).

All these works will soon become possible in great detail as
large-scale parallel supercomputers will enable ever more realistic
simulations of structure formation as part of the early universe, and
on large scales down to z=0. On a similar timescale increasingly
detailed observations of stars in the halo and ultra-faint dwarfs will
become available (e.g., Skymapper). However, observations alone will
likely not be able to uncover the underlying physical processes to
conclusively confirm the details of the assembly history of the Milky
Way stellar halo and the cosmological origin of the ancient
$\sim13$\,Gyr old stars which must have formed long before the Milky
Way was fully assembled.

\begin{acknowledgement}
A.F. gratefully acknowledges a Clay Fellowship which is administered
by the Smithsonian Astrophysical Observatory.
\end{acknowledgement}

\end{document}